\pdfoutput=1
\documentclass[12pt,titlepage]{article}

\font\grande=cmr9.5 scaled \magstep4
\font\medio=cmr9.5 scaled \magstep2
\outer\def\beginsection#1\par{\medbreak\bigskip
      \message{#1}\leftline{\bf#1}\nobreak\medskip
\vskip-\parskip
      \noindent}
\usepackage{graphicx} 
\begin{document}
\bibliographystyle {unsrt}

\titlepage

\begin{flushright}
\end{flushright}

\vspace{1cm}
\begin{center}
{\grande Post-inflationary phases stiffer than radiation}\\
\vspace{0.5cm}
{\grande and Palatini formulation}\\
\vspace{1cm}
 Massimo Giovannini 
 \footnote{Electronic address: massimo.giovannini@cern.ch} \\
\vspace{1cm}
{{\sl Department of Physics, CERN, 1211 Geneva 23, Switzerland }}\\
\vspace{0.5cm}
{{\sl INFN, Section of Milan-Bicocca, 20126 Milan, Italy}}
\vspace*{1cm}
\end{center}

\vskip 0.3cm
\centerline{\medio  Abstract}
\vskip 0.1cm
If the inflaton and the quintessence fields are identified, the background geometry evolves through a stiff epoch 
undershooting the expansion rate of a radiation-dominated plasma. For some classes of inflationary potentials this scenario is at odds with the current observational evidence since the corresponding tensor-to-scalar ratio is too large. Quintessential inflation is analyzed when the gravitational action is supplemented by a contribution quadratic in the Einstein-Hilbert term. In the Palatini formulation the addition such a term does not affect the scalar modes during the inflationary phase and throughout the course of the subsequent stiff epoch but it suppresses the tensor power spectrum and the tensor-to-scalar ratio. While in the Palatini formulation the power-law potentials leading to a quintessential inflationary dynamics are again viable, the high-frequency spike of the relic graviton spectrum is squeezed and the whole signal is suppressed at least when the higher-order contributions appearing in the action are explicitly decoupled from the inflaton.
\noindent

\vspace{5mm}
\vfill
\newpage

\renewcommand{\theequation}{1.\arabic{equation}}
\setcounter{equation}{0}
\section{Introduction}
\label{sec1}
The earliest direct test of the thermodynamic history of the universe relies on big-bang nucleosynthesis which is ultimately responsible for the formation of the light nuclear elements at an approximate temperature of the order of $0.1$ MeV. 
Prior thereto the expansion rate could have been very different from 
radiation and the first speculations along this direction date back to Zeldovich \cite{zel}, Sakharov \cite{sak} and Grishchuk \cite{gris}. After the formulation of conventional inflationary models Ford \cite{ford} 
noted that gravitational particle production at the end of inflation could account for the entropy of the present universe and observed that the backreaction effects of the created quanta constrain the length of a stiff post-inflationary phase by making the expansion dominated by radiation. It has been later argued by  Spokoiny \cite{spo} that various classes of scalar field potentials exhibit a transition from inflation to a stiff phase dominated by the kinetic energy of the inflaton; the same author also analyzed the conditions under which the expansion is again dominated by the inflaton asymptotically in the future. In more recent times it became increasingly plausible to have a single scalar field acting as inflaton in the early universe and as quintessence field in the late universe \cite{pee,stein}. A generic signature of a post-inflationary phase stiffer than radiation is the production of relic gravitons with increasing spectral energy density \cite{mg1}. 

In quintessential inflationary models the inflaton and the quintessence field are identified in a single scalar 
degree of freedom \cite{peevil,mg2} and various concrete forms of the inflaton-quintessence potential $V(\varphi)$ 
have been proposed and scrutinized through the years. The transition between an inflationary phase 
and a kinetic phase can be realized both with power-law potentials and with exponential potentials.  For instance the ``dual'' potentials evolve as a power-law during inflation and as an inverse power-law during the quintessential phase.  Probably the simplest example along this direction is given by $V(\varphi) = \lambda (\varphi^4 + M^4)$ 
for $\varphi < 0$ and $V(\varphi) = \lambda M^8/(\varphi^4 + M^4)$ for $\varphi \geq 0$ \cite{peevil,mg2}.
Modulated exponential potentials lead to similar dynamical evolutions \cite{spo} and the 
presence of a long stiff post-inflationary phase increases the maximal number of $e$-folds today 
accessible by large-scale observations \cite{liddle,zao,mg3}. While  the actual occurrence of a stiff epoch might not be related to conventional power-law inflation, it 
is disappointing that some of the simplest version of quintessential inflationary models are 
excluded from the current large-scale observations \cite{RT}.  For instance it is well known that the 
quartic potential leads to a scalar spectral index $n_{s}$ and to a tensor-to-scalar ratio $r_{T}$ that are excluded by the marginalized joint contour at $2\sigma$ confidence-level \cite{RT2}. A possible way out would be that since in quintessential inflation the number of e-folds is larger than in standard inflation, a quadratic (rather than quartic) potential leads to theoretical values in the $(n_{s} , r_{T} )$  plane that fall within the $2\sigma$ contours \cite{haro}.  

Within the Palatini approach it has been recently suggested that  in the presence of a generalized gravitational action, the slow-roll parameters and the tensor-to-scalar ratio can be suppressed in comparison with the conventional situation \cite{anto1,enqv,anto2,tenka}. For this purpose, instead of working in the framework of Einstein-Hilbert gravity (where the Palatini formulation just implies the metricity condition) we shall consider a generalized gravitational action containing higher-order terms.

Through the years the gravitational actions containing more than two derivatives of the metric have been explored  in radically diverse contexts. For instance the addition of higher-order curvature terms to the Einstein-Hilbert action are a key element of quantum theories in curved background geometries (see e.g. \cite{BD,PT} for two classic monographs on this theme).  In the early universe these terms can dominate and may lead to a large class of models where 
the scalar field driving inflation is in principle absent (see e.g. \cite{star1,noj}) but it reappears in a suitable 
conformally related frame where the typical inflationary potential is characterized by a 
quasi-flat plateau for large values of the putative scalar field. In string theory higher derivatives appear in the first 
string tension correction to the (tree-level) effective action  
\cite{des1,ts1,cal} and naturally arise at even higher-orders. In this context 
the Euler-Gauss-Bonnet combination plays a particular role: in four dimensions 
it corresponds to the Euler invariant whereas,  
in  dimensions larger than four, it does not lead to terms containing 
more than two derivatives of the metric with respect to the 
space-time coordinates \cite{mad1,mg2a}. Finally, if inflation is regarded as an effective 
theory, higher-order curvature corrections naturally appear when all the terms 
containing at least four derivatives are simultaneously considered in the effective 
inflationary action \cite{w1,od2}.

The common feature of the diverse models mentioned in the previous paragraph is that the higher derivatives terms in the action are always treated within the metric approach. However, when the gravitational action contains higher-order terms in the Ricci scalar, the Palatini and the metric formulations are notoriously inequivalent \cite{ST,HG}. This is part of the motivation of some recent analyses \cite{anto1,enqv,anto2,tenka} (see also \cite{maeda}) stressing that, unlike in the metric approach, in the Palatini formulation the $f(\overline{R})$ gravity does not introduce further degrees of freedom but just changes the relations between the existing ones. For the present purposes the general form can then be written as\footnote{We shall be using conventions where the four-dimensional metric has a signature mostly minus, i.e. $(+, \,-,\,-,\, -)$; Greek indices will run from $0$ to $3$ while Roman (lowercase) indices will refer to the three spatial coordinates. }:
\begin{equation}
S = - \frac{1}{2 \ell_{P}^2} \int d^{4} x\,\,\sqrt{-g} \,f(\overline{R})  + \int d^{4} x\,\ \sqrt{-g}\,\, \biggl[ \frac{1}{2} g^{\alpha\beta} \partial_{\alpha} \varphi \partial_{\beta} \varphi - V(\varphi) \biggr],
\label{one} 
\end{equation}
where $\ell_{P} = \sqrt{8 \pi G}$ and $\overline{R}$ denotes, for convenience, the Ricci scalar 
defined in terms of the Palatini connections $\overline{\Gamma}_{\alpha\beta}^{\,\,\,\lambda}$. 
Note that, within the present notations, $\ell_{P} = 1/\overline{M}_{P} = \sqrt{8\pi}/M_{P}$ and 
$M_{P} = 1.22\times 10^{19}$ GeV; both $M_{P}$ and $\overline{M}_{P}$ shall be employed hereunder when needed.  To avoid ambiguities it must be stressed that, within the notations of Eq. (\ref{one}),  the Riemann tensor is defined in terms of the Palatini connections:
\begin{equation}
\overline{R}^{\lambda}_{\,\,\,\,\alpha\nu\beta} = \partial_{\nu} \overline{\Gamma}_{\alpha\beta}^{\,\,\,\,\,\,\lambda} - \partial_{\beta} \overline{\Gamma}_{\nu\alpha}^{\,\,\,\,\,\,\lambda}  + \overline{\Gamma}_{\alpha\beta}^{\,\,\,\,\,\,\delta}\, \overline{\Gamma}_{\nu\delta}^{\,\,\,\,\,\,\lambda} -  \overline{\Gamma}_{\nu\alpha}^{\,\,\,\,\,\,\delta}\, \overline{\Gamma}_{\beta\delta}^{\,\,\,\,\,\,\lambda}.
\label{two}
\end{equation}
 The Ricci tensor follows from Eq. (\ref{two}) as $\overline{R}_{\alpha\beta} = \overline{R}^{\lambda}_{\,\,\,\,\alpha\lambda\beta}$. Finally $\overline{\nabla}_{\alpha}$ and $\nabla_{\alpha}$ will correspond to the covariant derivatives constructed from the Palatini connections and from the standard Christoffel symbols respectively\footnote{We shall often use the Levi-Civita connections as a synonym for the standard 
 Christoffel symbols even if the two notions are slightly different from a purely geometrical viewpoint.}. Therefore, for instance, the variation of the Riemann tensor relevant to the Palatini formulation will be $\delta \overline{R}^{\lambda}_{\,\,\,\,\alpha\nu\beta} = \overline{\nabla}_{\nu} \overline{\Gamma}_{\alpha\beta}^{\,\,\,\,\,\,\lambda} - \overline{\nabla}_{\beta} \overline{\Gamma}_{\nu\alpha}^{\,\,\,\,\,\,\lambda}$ where the covariant derivatives are not defined in terms of the conventional Christoffel connections. 
 
If we assume that $V(\varphi)$ defines a quintessential inflationary potential when the gravitational action has the standard Einstein-Hilbert form (i.e. $f(\overline{R}) =\overline{R}$ in the notations of Eq. (\ref{one})) what happens when higher-order corrections are consistently included? What will be the evolution of the scalar and tensor modes of the geometry during inflation and in the subsequent stiff phases? To address these questions it will be important to deal with the explicit evolution of the large-scale inhomogeneities induced by the action (\ref{one}). A detailed analysis, generally relevant for the Palatini formulation, has been presented in Ref. \cite{pert1}. We shall however follow a different approach based on the evolution in the Einstein frame. As we shall specifically argue the evolution of the background and of the fluctuations in the two frames will be ultimately the same as it happens in the case of the metric formulation \cite{mg4} so that the most convenient frame can always be exploited. It will be finally useful to approach the evolution of the inhomogeneities in reasonably general terms so that it will be  applicable both during inflation and all along the stiff epoch. In the discussion of the inhomogeneities the potential will be kept to be generic. However, for illustration, it will be useful to consider the possibility of a dual potential interpolating between inflationary and quintessential dynamics.

The layout of this investigation is  the following. In section \ref{sec2} we shall analyze the evolution of Eq. (\ref{one}) in two different conformally related frames (i.e. the Palatini and the Einstein frames) by exploiting the corresponding equations of motion; at the end we shall also discuss the effective action of the system. In section \ref{sec3} the evolution of the background will be studied both during inflation and in the subsequent stiff epoch. Section \ref{sec4} is devoted to the power spectra of the scalar and tensor modes of the geometry; the suppression of the tensor to scalar ratio and the general evolution equations will be explicitly addressed. The same combination governing the perturbative  corrections of the background evolution also enter the evolution of the curvature  perturbations on comoving orthogonal hypersurfaces. Section \ref{sec5} contains the concluding remarks. 
\renewcommand{\theequation}{2.\arabic{equation}}
\setcounter{equation}{0}
\section{The dynamics in different frames}
\label{sec2}
The extremization of the action (\ref{one}) with respect to the variation of the Palatini connections 
$\overline{\Gamma}_{\alpha\beta}^{\,\,\,\,\,\,\lambda}$ implies the condition
\begin{equation}
\overline{\nabla}_{\lambda} \biggl[ \sqrt{- g} \,\, g^{\alpha\beta} F \biggr] =0 ,
\qquad F = \frac{\partial f}{\partial \overline{R}},
\label{FF1}
\end{equation}
where $F$ denotes throughout the derivative of $f$ with respect to $\overline{R}$. 
Equation (\ref{FF1}) does not define a Levi-Civita connection however it can be brought in that 
form by defining an appropriately rescaled metric $\overline{g}_{\alpha\beta} = \Omega^2 
g_{\alpha\beta}$. Provided the conformal factor is appropriately chosen (i.e. $\Omega = \sqrt{ F}$ in four space-time dimensions), Eq. (\ref{FF1}) becomes $\overline{\nabla}_{\lambda} [\sqrt{- \overline{g}} \,\, \overline{g}^{\alpha\beta}]=0$ which now implies that the Palatini connection does have the Levi-Civita form in terms of the 
rescaled metric $\overline{g}_{\alpha\beta}$ but {\em not} in terms of the original $g_{\alpha\beta}$. 
The two different metrics $\overline{g}_{\alpha\beta}$ and $g_{\alpha\beta}$ define 
two complementary physical descriptions that will be generically referred to as the Einstein and the Palatini 
frame respectively. While the inflationary evolution is more conveniently studied  
in the Einstein frame, the stiff dynamics becomes simpler in the Palatini frame. The two 
frames will be shown to be equivalent both at the level of the background and for the 
corresponding inhomogeneities. 

Note that $\overline{R}(\overline{g}) = \overline{g}^{\alpha\beta} \overline{R}_{\alpha\beta}$ 
denotes the Ricci scalar defined by contraction with the Einstein frame metric $\overline{g}^{\alpha\beta}$.
In the present context, however, it is also useful to introduce the Ricci scalar defined by contraction 
with the Palatini metric, i.e. $\overline{R}(g) = g^{\alpha\beta} \overline{R}_{\alpha\beta}$.
For instance the action (\ref{one}) is written in terms of $\overline{R}_{\alpha\beta}$ but the metric 
used for the contraction is $g^{\alpha\beta}$ (and not $\overline{g}^{\alpha\beta}$).  The extremization of the action (\ref{one}) with respect to the variation of the metric $g^{\mu\nu}$ implies the validity of the following equation:
\begin{eqnarray}
&& F \, \overline{R}_{\mu\nu}(g) - \frac{f}{2} \, g_{\mu\nu} = \ell_{P}^2 \, T_{\mu\nu}(g), 
\label{FF1a}\\
&& T_{\mu\nu}(g) = \partial_{\mu} \varphi \partial_{\nu} \varphi - g_{\mu\nu} \biggl(\frac{1}{2} g^{\alpha\beta} \partial_{\alpha} \varphi \partial_{\beta} \varphi - V\biggr),
\label{FF1aa}
\end{eqnarray}
where the energy-momentum tensor is defined in the Palatini frame.
If we now contract Eq. (\ref{FF1a}) with the $g^{\mu\nu}$ we obtain the explicit relation 
between $\overline{R}(g)$ and $T(g)$:
\begin{equation}
F\, \overline{R}(g) - 2 \, f = \ell_{P}^2\, T(g).
\label{FF1ab}
\end{equation} 
Equation (\ref{FF1ab}) leads to a complicated relation that depends on the analytic
expression of $f(\overline{R})$; various specific forms of  $f(\overline{R})$ have been 
concocted in the past for different purposes and, in particular, to model the evolution of dark energy during the recent stages of evolution of the background after the onset of the matter-dominated epoch (see e.g.  \cite{rr1,rr2}).  In the present context the problem is different since 
the form of $f(\overline{R})$ must only delicately affect the scalar and tensor 
modes of the geometry that will be discussed later on in section \ref{sec4}. 

An interesting possibility analyzed in \cite{anto1,enqv,anto2,tenka} 
stipulates that $f(\overline{R})$ could have a quadratic form implying, according to Eq. (\ref{FF1ab}), a linear relation between $\overline{R}$ and $T(g)$. Consistently with the present conventions on the signature of the metric we shall therefore choose
 $f(\overline{R}) = \overline{R} - \overline{\alpha} \overline{R}^2 $ where $\overline{\alpha} = \alpha/M^2$ and 
$M \leq \overline{M}_{P}$ denotes  a typical mass scale; we also have, as anticipated, that in the quadratic case
Eq. (\ref{FF1ab}) generically implies  $\overline{R}(g) = - \ell_{P}^2 \,T(g)$ so that the concrete form of $F$ becomes:
\begin{equation} 
F = 1 + 2 \, \overline{\alpha} \, \ell_{P}^2 \, \biggl[ 4 V - g^{\alpha\beta} \partial_{\alpha} \varphi \partial_{\beta} \varphi\biggr].
\label{FF1b}
\end{equation}
Equation (\ref{FF1b}) holds in the Palatini frame; since in the Einstein frame  $g^{\alpha\beta} = F\,\, \overline{g}^{\alpha\beta}$,
Eq. (\ref{FF1b}) becomes:
\begin{equation} 
F = 1 + 2 \, \overline{\alpha} \, \ell_{P}^2 \,\biggl[ 4 V - \overline{g}^{\alpha\beta} \, F\, \partial_{\alpha} \varphi \partial_{\beta} \varphi\biggr].
\label{FF1c}
\end{equation}
For the quadratic case (and in the Einstein frame) $f$ can be directly expressed in terms of $F$, namely
\begin{equation} 
f = \frac{\ell_{P}^2}{2} \, (F+1) \,\biggl[ 4 V - \overline{g}^{\alpha\beta} \, F\, \partial_{\alpha} \varphi \partial_{\beta} \varphi\biggr].
\label{FF1e}
\end{equation}
While in the Palatini frame Eq. (\ref{FF1b}) defines directly $F$, in the Einstein frame Eq. (\ref{FF1c}) becomes an algebraic 
condition that can be eventually solved and the result is:
\begin{equation}
F = \frac{F_{0}}{F_{1}}, \qquad F_{0} = 1 + 8\, \overline{\alpha}\,  \ell_{P}^2\, V, \qquad F_{1} = 1 + 2 \, \overline{\alpha} \, \ell_{P}^2 \, \overline{g}^{\alpha\beta} \, \partial_{\alpha} \varphi \partial_{\beta} \varphi.
\label{FF1d}
\end{equation}
Equations (\ref{FF1c}) and (\ref{FF1d})--(\ref{FF1e}) will be quite relevant when discussing the equivalence 
between the Palatini and the Einstein frame.

\subsection{The Palatini frame}
The field equations in the Palatini frame can also be expressed in a more conventional form 
 by introducing the corresponding Einstein tensor:
\begin{eqnarray}
&& \overline{G}_{\mu\nu} + \frac{f}{2F} g_{\mu\nu} = \frac{\ell_{P}^2}{F} \biggl[ T_{\mu\nu}(g) - \frac{1}{2} g_{\mu\nu} T(g) \biggr],
\label{FF2}\\
&& g^{\alpha\beta} \nabla_{\alpha} \nabla_{\beta} \varphi + \frac{\partial V}{\partial \varphi} =0,
\label{FF3}
\end{eqnarray}
where Eq. (\ref{FF2}) follows from Eqs. (\ref{FF1a}) and (\ref{FF1aa}) while Eq. (\ref{FF3}) 
come from the extremization of the action (\ref{one}) with respect to the variation of $\varphi$.
To obtain an explicit form of the equations of motion, the Einstein tensor $\overline{G}_{\mu\nu}$ must be expressed in terms of the metric $g_{\mu\nu}$; the transformation reads:
\begin{eqnarray}
\overline{R}_{\mu\nu} &=& R_{\mu\nu} + 2 \biggl[ \partial_{\mu} q \partial_{\nu} q - g_{\mu\nu} (\partial q)^2\biggr] - 2 \biggl[ \nabla_{\mu} \nabla_{\nu} q + \frac{g_{\mu\nu}}{2} \nabla^2 q \biggr], 
\label{FF5}\\
\overline{R} &=& R - 6 \nabla^2q - 6 (\partial q)^2 ,
\label{FF6}\\
\overline{G}_{\mu\nu} &=& G_{\mu\nu} +  2 \biggl[\partial_{\mu} q \partial_{\nu} q + \frac{g_{\mu\nu}}{2}(\partial q)^2\biggr]
- 2 \biggl[ \nabla_{\mu} \nabla_{\nu} q -   g_{\mu\nu} \nabla^2 q\biggr],
\label{FF7}
\end{eqnarray}
where $q$ is the natural logarithm of the conformal factor (i.e.  $q = \ln{\sqrt{F}}$); moreover, for 
notational convenience the shorthand notations $\nabla^2 q = g^{\alpha\beta} \nabla_{\alpha} \nabla_{\beta} q$ and $(\partial q)^2 = g^{\alpha\beta} \partial_{\alpha} q \partial_{\beta} q$ will be used throughout. Inserting Eq. (\ref{FF7}) into Eq. (\ref{FF2}) we obtain 
\begin{eqnarray}
G_{\mu\nu} + \biggl[\frac{f}{2 F} + 2 \nabla^2 q  + (\partial q)^2\biggr] g_{\mu\nu} &=& \frac{\ell_{P}^2}{F} \biggl[ \partial_{\mu} \varphi \partial_{\nu} \varphi - V g_{\mu\nu} \biggr] +  2 \biggl[ \nabla_{\mu} \nabla_{\nu} q - \partial_{\mu} q \partial_{\nu} q\biggr].
\label{FF8}
\end{eqnarray}
The approach based on Eqs. (\ref{FF2}), (\ref{FF3}) and (\ref{FF8}) is convenient since 
the equation of $\varphi$ which is standard. The price to pay is that the evolution of the metric 
is complicated by the presence of $q$ and of its derivatives. Note that Eq. (\ref{FF8}) is still general since the form of $f$ and $F$ has not been specified. 

\subsection{The Einstein frame}
The evolution in the Einstein frame follows by using the conformal rescaling suggested in Eq. (\ref{FF1})
and  the evolution equations of the system can be written in the following form:
\begin{eqnarray}
&& \overline{G}_{\mu\nu} + \frac{f}{2 F^2} \overline{g}_{\mu\nu} = \frac{\ell_{P}^2}{F} \biggl[ \partial_{\mu} \varphi \partial_{\nu} \varphi - \biggl(\frac{V}{F}\biggr) \overline{g}_{\mu\nu} \biggr],
\label{FF9}\\
&& \overline{g}^{\mu\nu} \overline{\nabla}_{\mu} \overline{\nabla}_{\nu} \varphi + \frac{1}{F} \frac{\partial V}{\partial \varphi}= 2 \overline{g}^{\alpha\beta} \partial_{\alpha} q \partial_{\beta} \varphi.
\label{FF10}
\end{eqnarray}
Note that while $\nabla_{\alpha} \varphi = \overline{\nabla}_{\alpha} \varphi$ we clearly have that 
$\overline{\nabla}_{\mu} \overline{\nabla}_{\nu} \varphi \neq \nabla_{\mu} \nabla_{\nu} \varphi$. 
Thanks to Eq. (\ref{FF1e}),  Eq. (\ref{FF9}) can be further simplified in the quadratic case and the result is:
\begin{equation}
\overline{G}_{\mu\nu} = \frac{\ell_{P}^2}{F} \biggl\{ \partial_{\mu} \varphi \partial_{\nu} \varphi + 
\overline{g}_{\mu\nu}\biggl[ V - \frac{(F+1)}{4} \overline{g}^{\alpha\beta} \partial_{\alpha} \varphi \partial_{\beta} \varphi\biggr] \biggr\}.
\label{FF11}
\end{equation}
Recalling finally Eq. (\ref{FF1d}), the expression of Eq. (\ref{FF11}) 
becomes even more explicit: 
\begin{equation}
\overline{G}_{\mu\nu} = \ell_{P}^2 \biggl\{ F_{1} \partial_{\mu} \overline{\varphi}\, \partial_{\nu} \overline{\varphi} +
\overline{g}_{\mu\nu} \biggl[ F_{1} W - \frac{(F_{0} + F_{1})}{4}\, \overline{g}^{\alpha\beta} \partial_{\alpha} \overline{\varphi} \,
\partial_{\beta} \overline{\varphi} \biggr] \biggr\}.
\label{FF12}
\end{equation}
The kinetic term of the scalar field appearing in Eq. (\ref{FF12}) and the corresponding potential can be rescaled through $F_{0}$ according to the following transformation:
\begin{equation}
\partial_{\mu} \varphi \to \partial_{\mu} \overline{\varphi} = \frac{\partial_{\mu} \varphi}{\sqrt{F_{0}}}, \qquad V \to W= \frac{V}{F_{0}}.
\label{FF13}
\end{equation}
The rescaling of Eq. (\ref{FF13}) does not involve any supplementary degree of freedom since $F_{0}$ solely depends 
on the scalar field potential. While this rescaling is not strictly necessary, it is convenient to make a
more direct contact with the current literature. Consequently, thanks to Eq. (\ref{FF13}) the result of Eq. (\ref{FF12}) can be finally recast in the following form:
\begin{eqnarray}
\overline{G}_{\mu\nu} &=& \ell_{P}^2 \biggl\{ ( 1 + 2 Q ) \partial_{\mu} \overline{\varphi}\, \partial_{\nu} \overline{\varphi}  
- \overline{g}_{\mu\nu} \biggl[ W - \frac{(\partial \overline{\varphi})^2}{2} ( 1 + Q)\biggr] \biggr\},
\nonumber\\
Q &=& 1 + 2 \overline{\alpha} \, \ell_{P}^2 \, F_{0} \, (\partial \overline{\varphi})^2, \qquad (\partial \overline{\varphi})^2 = \overline{g}^{\alpha\beta} 
\partial_{\alpha} \overline{\varphi} \,\partial_{\beta} \overline{\varphi}.
\label{FF14}
\end{eqnarray}
With the same strategy and the rescaling of Eq. (\ref{FF13}), Eq. (\ref{FF10}) can also be written in a similar manner:
\begin{equation}
\frac{1}{\sqrt{-\overline{g}}} \partial_{\alpha} \biggl[ \sqrt{-\overline{g}}\,\, \overline{g}^{\alpha\beta}\, \, \partial_{\beta} \overline{\varphi} \,\,( 1 + 2 Q) \biggr] + (1 - 4 Q^2 )\frac{\partial W}{\partial \overline{\varphi} } =0.
\label{FF15a}
\end{equation}
While the discussion  has been conducted so far in terms of the equations of motion, 
the same results can be derived by modifying the action in 
the Einstein frame, as it will be shown hereunder.

\subsection{The viewpoint of the action}

The same results obtained by modifying the equations of motion follow by transforming the action (\ref{one}) 
from the Palatini to the Einstein frame. To clarify this point 
Eq. (\ref{one}) shall be written as: 
\begin{equation}
S =  -\frac{1}{2 \ell_{P}^2} \int d^{4} x \sqrt{-g} \biggl[ f(\lambda) + F(\lambda) ( \overline{R} - \lambda) \biggr] +
\int d^{4} x \sqrt{-g} \biggl[ \frac{1}{2} \overline{g}^{\alpha\beta} \partial_{\alpha} \varphi \partial_{\beta} \varphi - V(\varphi)\biggr]
\label{FF15}
\end{equation}
where $\lambda$ now represents an appropriate Lagrange multiplier while, by definition, $F(\lambda) =\partial_{\lambda} f$. By extremizing the action with respect to the variation of $\lambda$ we obtain the condition
$\lambda = \overline{R}$ and the action (\ref{one}) is recovered provided $\partial_{\lambda} F \neq 0$.
If the metric $g_{\mu\nu}$ is conformally rescaled as suggested by Eq. (\ref{FF1}), the action of Eq. (\ref{FF15}) assumes the form:
\begin{equation}
S = - \frac{1}{2 \ell_{P}^2} \int d^{4} x \sqrt{- \overline{g}} \biggl[ \overline{R}(\overline{g}) - \frac{\lambda F(\lambda) - f(\lambda)}{F^2} \biggr]
+ \int d^4 x \sqrt{- \overline{g}} \biggl[ \frac{1}{2 F} \overline{g}^{\alpha\beta} \partial_{\alpha} \varphi \partial_{\beta} \varphi - \frac{V}{F^2}\biggr],
\label{FF16}
\end{equation}
where now the Einstein-Hilbert term is standard (i.e. formally decoupled from $f(\lambda)$ and its derivatives).
Assuming then a quadratic form (i.e.  $f(\lambda) = \lambda - \overline{\alpha} \lambda^2$) we have $F(\lambda) = 1 - 2 \overline{\alpha} \lambda$
which also implies $\lambda = (1 - F)/(2 \overline{\alpha})$. If this result is substituted back into Eq. (\ref{FF16}) the action becomes:
\begin{equation}
S = - \frac{1}{2 \ell_{P}^2} \int d^{4} x \sqrt{- \overline{g}}\, \biggl[ \overline{R}(\overline{g}) + \frac{(F -1)^2}{4 \overline{\alpha} F^2} \biggr]
+ \int d^4 x \sqrt{- \overline{g}} \biggl[ \frac{1}{2 F} \overline{g}^{\alpha\beta} \partial_{\alpha} \varphi \partial_{\beta} \varphi - \frac{V}{F^2}\biggr].
\label{FF17}
\end{equation}
The extremization of Eq. (\ref{FF17}) respect to $\delta F$ implies exactly the same condition already 
discussed in Eq. (\ref{FF1d}). Therefore inserting back into Eq. (\ref{FF17}) the condition Eq. (\ref{FF1d}) (and using the notation 
$F= F_{0}/F_1$) the action takes this final form:
\begin{equation}
S= - \frac{1}{2 \ell_{P}^2} \int d^{4} x \sqrt{- \overline{g}} \,\,\overline{R}(\overline{g}) + 
 \int d^{4} x \sqrt{- \overline{g}} \biggl\{ \frac{1}{2} \overline{g}^{\alpha\beta} \partial_{\alpha} \overline{\varphi} \partial_{\beta} \overline{\varphi} \biggl[ 1 + Q \biggr]- W\biggl\}, 
\label{FF18}
\end{equation}
where, as in Eq. (\ref{FF14}), $ Q=\overline{\alpha} \,\ell_{P}^2 \, F_{0} \, \overline{g}^{\mu\nu} \, \partial_{\mu} \overline{\varphi}\, \partial_{\nu} \overline{\varphi}$.  Equations (\ref{FF14}) and (\ref{FF15a}) can now be obtained 
again by extremizing the action with respect to the variation of $\overline{g}^{\mu\nu}$ and $\overline{\varphi}$. Since $Q$ implicitly depends on $W$ and $\overline{\varphi}$,  its variation 
must be appropriately included in the final expression. Equation (\ref{FF18}) is anyway revealing since it demonstrates that, as long as $Q$ is negligible, 
the standard scalar-tensor theory is recovered and the inflaton is minimally coupled to the geometry.

\renewcommand{\theequation}{3.\arabic{equation}}
\setcounter{equation}{0}
\section{Background, slow-roll and stiff evolution}
\label{sec3}
\subsection{General evolution of the background}
In a conformally flat metric of Friedmann-Robertson-Walker type $\overline{g}_{\mu\nu} = \overline{a}^2(\tau) \eta_{\mu\nu}$ (where $\eta_{\mu\nu}$ denotes the Minkowski metric) 
the evolution equations in the Einstein frame follow from Eqs. (\ref{FF9}) and (\ref{FF10}):
\begin{eqnarray}
&&\overline{\mathcal H}^2 = \frac{\ell_{P}^2}{3 F} \biggl[ \frac{(3 - F)}{4} \varphi^{\prime\,\, 2} + V\, \overline{a}^2\biggr],
\label{SS1}\\
&& (\overline{\mathcal H}^2 + 2 \overline{{\mathcal H}}^{\,\prime}) =  \frac{\ell_{P}^2}{F} \biggl[ V \, \overline{a}^2 - \frac{(1 + F)}{4}  \varphi^{\prime\,\, 2}\biggr],
\label{SS2}\\
&& \varphi^{\prime\prime} + 2\, \overline{{\mathcal H}} \,\varphi^{\prime} + \frac{\overline{a}^2}{F} \frac{\partial V}{\partial \varphi} = 2\, q^{\prime} \,\varphi^{\prime},
\label{SS3}
\end{eqnarray}
where the prime denotes a derivation with respect to the conformal time coordinate while $\overline{{\mathcal H}}$.
It is relevant to mention that the coformal time coordinate $\tau$ (unlike the cosmic time coordinate) does not change from the Einstein to the Palatini frame. Therefore the prime will consistently denote a derivation 
with respect to $\tau$ in both frames.
The potential term can be eliminated between Eqs. (\ref{SS1}) and (\ref{SS3}) and the resulting combination 
reads 
\begin{equation}
\overline{\mathcal H}^2 - \overline{\mathcal H}^{\,\prime} = \frac{\ell_{P}^2}{2 F} \varphi^{\prime\,\, 2}.
\label{SS4}
\end{equation}
The form of Eqs. (\ref{SS1})--(\ref{SS3}) and (\ref{SS4}) in the Palatini 
frame follows from Eqs. (\ref{FF2})--(\ref{FF3}); a swifter (and equivalent) derivation 
follows  by applying the conformal transformation in its concrete background expression
\begin{equation}
\overline{a} \to \overline{a} = \sqrt{F} a, \qquad \overline{{\mathcal H}} = {\mathcal H} + q^{\prime},
\label{SS5}
\end{equation}
directly to Eqs. (\ref{SS1})--(\ref{SS3}) and (\ref{SS4}). 
For instance, according to Eq. (\ref{SS5}), if we transform Eqs. (\ref{SS3}) and (\ref{SS4}) 
we immediately obtain 
\begin{eqnarray}
&& \varphi^{\prime\prime} + 2 {\mathcal H} \varphi^{\prime} + a^2\frac{\partial V}{\partial \varphi} =0,
\label{SS6}\\
&& {\mathcal H}^2 - {\mathcal H}^{\prime} = \frac{\ell_{P}^2}{2 F} \varphi^{\prime\,\, 2} + q^{\prime\prime} - 2 {\mathcal H} q^{\prime} - q^{\prime\,\, 2},
\label{SS7}
\end{eqnarray}
and similarly for Eqs. (\ref{SS1}) and (\ref{SS2}). Equations (\ref{SS6}) and (\ref{SS7}) coincide 
with the expressions following directly  from Eqs. (\ref{FF2})--(\ref{FF3})  written in the conformally rescaled frame $g_{\mu\nu} = a^2(\tau) \eta_{\mu\nu}$.  The equivalence between the Palatini and Einstein frame illustrated above implies that if and when the dynamics is solved in one frame, the evolution in the conformally related frame can be obtained by applying the background transformation of Eq. (\ref{SS5}). This equivalence is also preserved at the level of the gauge-invariant fluctuations,
as we shall see in section \ref{sec4}. 

Once the form of $F$ is fixed, Eqs, (\ref{SS1})--(\ref{SS3}) and (\ref{SS4}) can be further modified. 
In particular, recalling the choice of Eq. (\ref{FF1d}) we have that the background becomes
\begin{equation}
\overline{F} = \frac{\overline{F}_{1}}{\overline{F}_{0}}, \qquad \overline{F}_{1} = 1 + 2 \overline{\alpha}\, \ell_{P}^2 \,\frac{\varphi^{\prime\,\, 2}}{\overline{a}^2}, \qquad \overline{F}_{0} = 1 + 8 \,\overline{\alpha} \,\ell_{P}^2 \,V,
\label{SS8}
\end{equation}
where the overline now distinguishes the general expression of $F$ from its background value.
Inserting Eq. (\ref{SS8}) into Eqs.  (\ref{SS1})--(\ref{SS3}) and (\ref{SS4}) 
the explicit form of the background evolution becomes:
\begin{eqnarray}
&&\overline{\mathcal H}^2 = \frac{\ell_{P}^2}{3} \biggl[ \frac{\overline{\varphi}^{\,\prime\,2}}{2} ( 1 + 3 \overline{Q}) + W\, \overline{a}^2\biggr],
\label{SS9}\\
&& (\overline{\mathcal H}^2 + 2 \overline{{\mathcal H}}^{\,\prime}) =  \ell_{P}^2 \biggl[ W \, \overline{a}^2 - \frac{\varphi^{\,\prime\,2}}{2} ( 1 + \overline{Q})\biggr],
\label{SS10}\\
&& \overline{\varphi}^{\prime\prime} + 2 \, \overline{{\mathcal H}} \,\overline{\varphi}^{\,\prime} + \frac{ 2 \overline{\varphi}^{\,\prime} \, \overline{Q}^{\,\prime}}{1 + 2 \overline{Q}} +\overline{a}^2 \,\frac{\partial W}{\partial \overline{\varphi}} =0. 
\label{SS11}
\end{eqnarray}
Equations (\ref{SS9}), (\ref{SS10}) and (\ref{SS11}) have been written in terms of the rescaled kinetic term and potential 
introduced in Eq. (\ref{SS5}); the explicit background expression for the rescalings of Eq. (\ref{SS5}) is given by: 
\begin{equation}
\overline{\varphi}^{\prime}= \frac{\varphi^{\prime}}{\sqrt{\overline{F}_{0}}}, \qquad W= \frac{V}{\overline{F}_{0}}.
\label{SS12}
\end{equation}
The potential term can be eliminated between Eqs. (\ref{SS1}) and (\ref{SS3}) and the resulting combination 
reads: 
\begin{equation}
\overline{\mathcal H}^2 - \overline{\mathcal H}^{\,\prime} = \frac{\ell_{P}^2}{2} \overline{\varphi}^{\,\prime\, 2} ( 1 + 2 \overline{Q}),
\label{SS13}
\end{equation}
which also follows directly from Eq. (\ref{SS4}). We also introduced, for convenience, the following shorthand notation
\begin{equation}
\overline{Q} = \frac{2 \, \overline{\alpha} \, \ell_{P}^2}{\overline{a}^2} \,\overline{\varphi}^{\,\prime\,2} \, \biggl[ 1 + 8\, \overline{\alpha}\, \ell_{P}^2 V\biggr],
\label{SS14}
\end{equation} 
that follows directly by evaluating Eq. (\ref{FF14}) on the background. Equation 
(\ref{SS14}) suggests the existence of two complementary regimes where we can be plausibly have that 
$\overline{Q} \ll 1$ in spite the value of $\alpha$. The first regime is the one where $\overline{\varphi}^{\,\prime\,2} \ll \overline{M}^2_{P} \, {\mathcal H}^2$ and it corresponds to a slow-roll evolution where in fact $\overline{\varphi}^{\,\prime\,2} \ll \overline{a}^2 \,V$.  The second complementary situation is the one where $\overline{a}^2 \, V \ll \overline{\varphi}^{\,\prime\,2}$ and it corresponds to a stiff evolution dominated by the kinetic energy. 
The same hierarchy controlling the dynamics of the background in the two aforementioned 
regimes also impact on the evolution of the corresponding inhomogeneities 
as it will be argued in section \ref{sec4}.

\subsection{Inflationary dynamics}
Instead of positing a specific potential, it is more instructive 
to scrutinize the impact of the the quadratic terms when $\alpha \neq 0$ by assuming that  inflation already occurs in the conventional situation (i.e. when $\alpha \to 0$). In the limit $\alpha \to 0$ we have that  $W \to V$, $\dot{\overline{\varphi}} \to \dot{\varphi}$ and the slow-roll parameters will then be given by: 
\begin{equation} 
\epsilon^{(0)} = \frac{\overline{M}_{P}^2}{2 V^2} \biggl(\frac{\partial V}{\partial \varphi} \biggr)^2\ll 1, \qquad 
\eta^{(0)} =  \frac{\overline{M}_{P}^2}{ V} \frac{\partial^2 V}{\partial \varphi^2} \ll 1.
\label{SS15}
\end{equation}
In the opposite limit (i.e. $\alpha \neq 0$), according to Eqs. (\ref{SS10})--(\ref{SS12})  the slow-roll 
dynamics follows from the approximated equations that we write, for convenience, in the cosmic 
time coordinate:
\begin{equation}
3 \overline{H}^{(\alpha)} \, \dot{\overline{\varphi}} + \frac{\partial W}{\partial \overline{\varphi}} =0, \qquad 
3 \overline{H}^{(\alpha)\, 2} \overline{M}_{P}^2 = W, \qquad 2 \overline{M}_{P}^2 \dot{\overline{H}}^{(\alpha)} = - \dot{\overline{\varphi}}^2,
\label{SS15a}
\end{equation} 
where $\overline{H}^{(\alpha)}$ denotes the Hubble rate; the superscript $(\alpha)$ reminds that Eqs. 
(\ref{SS15a}) refer to the case $\alpha \neq 0$ where the slow-roll parameters are defined as:
\begin{equation} 
\epsilon^{(\alpha)} = \frac{\overline{M}_{P}^2}{2 W^2} \biggl(\frac{\partial W}{\partial \overline{\varphi}} \biggr)^2, \qquad 
\eta^{(\alpha)} =  \frac{\overline{M}_{P}^2}{ W} \frac{\partial^2 W}{\partial \overline{\varphi}^2}.
\label{SS16}
\end{equation}
The explicit expressions of the slow-roll parameters may be obviously modified by using directly Eqs. (\ref{SS15a});
 so for instance we will have that $\epsilon^{(\alpha)} = - \dot{\overline{H}}^{(\alpha)}/\overline{H}^{(\alpha)\, 2}$.

We shall next consider the case when the corresponding slow-roll parameters are all small in the limit $\alpha \to 0$ (i.e. $\overline{F}_{0} \to 1$ and $\overline{F}_{1} \to 1$) and investigate what happens for a generic $\alpha \gg 1$. The relation between Eqs. (\ref{SS16}) and (\ref{SS15}) follows by recalling that $\partial \overline{\varphi} = \partial \varphi/\sqrt{F_{0}}$ and that  $W = V/F_{0}$; in particular, using the identity
\begin{equation}
\frac{1}{W} \biggl(\frac{\partial W}{\partial \overline{\varphi}} \biggr) = \biggl(\frac{\partial V}{\partial \varphi} \biggr) \biggl[ \frac{1}{F_{0}} - 
\frac{8 \overline{\alpha} \ell_{P}^2 V}{F_{0}^2} \biggr] \frac{F_{0}^{3/2}}{V} \equiv \frac{1}{ V \, \sqrt{F_{0}}}  \biggl(\frac{\partial V}{\partial \varphi} \biggr). 
\label{SS17}
\end{equation}
 and its analog (valid for the second derivatives of $W$ with respect to $\overline{\varphi}$), Eqs. (\ref{SS16}) and (\ref{SS15}) can be rewritten as:
\begin{eqnarray}
\epsilon^{(\alpha)} &=& \frac{\epsilon^{(0)}}{1 + 8 \,\overline{\alpha} \,\ell_{P}^2 \, V},
\label{SS18}\\
\eta^{(\alpha)} &=& \eta^{(0)} - \frac{24\, \epsilon^{(0)}\,\overline{\alpha}\, \ell_{P}^2\, V}{1 + 8 \, \overline{\alpha}\, \ell_{P}^2\, V}.
\label{SS19}
\end{eqnarray}
In the limit $\alpha \gg 1$ Eq. (\ref{SS19}) implies that  $\eta^{(\alpha)} = \eta^{(0)} - 3 \epsilon^{(0)}$ (recall 
$\alpha = M^2 \overline{\alpha}$); in the same limit, Eq. (\ref{SS18}) the value of the corresponding slow-roll parameter is rescaled in comparison with the conventional case when the quadratic terms are absent (i.e. when $\alpha \to 0$).

To derive Eqs. (\ref{SS18}) and (\ref{SS19}) we assumed that that $\overline{Q} \ll 1$ in Eqs. (\ref{SS9})--(\ref{SS11}): this requirement is always verified, even in the case $\alpha \gg 1$ since, by definition, 
\begin{equation}
\overline{Q} = \frac{4}{3} \overline{\alpha} \, \ell_{P}^2 \, \epsilon^{(\alpha)} V \equiv \frac{4 \epsilon^{(0)}}{3} \frac{\overline{\alpha} \ell_{P}^2 V}{1 + 8 \overline{\alpha} \,\ell_{P}^2 \,V},
\label{SS20}
\end{equation}
where the second equality follows immediately from Eq. (\ref{SS18}). All in all, provided $\epsilon^{(0)}\ll 1$ (as assumed in Eq. (\ref{SS15})) we will also have that $\overline{Q} \ll 1$ even if $\alpha \gg 1$. It actually 
follows from Eq. (\ref{SS20}) that when $\alpha \gg 1$ we will still have that $\overline{Q} \ll 1$ since  $\overline{Q} \to  \epsilon^{(0)}/6$ for $\alpha \gg 1$. Note that, in the limit $\alpha \gg 1$, the rescaled potential is suppressed as $W \to \overline{M}_{P}^2 M^2/(8 \alpha)$; indeed, by definition, $W = V/(1 + 8 \overline{\alpha} \ell_{P}^2 V)$ implying $ W \to 1/(8 \overline{\alpha} \ell_{P}^2)$. In spite of the shape of the potential, the terms quadratic in the Einstein-Hilbert action produce a suppression of the effective potential in comparison with the case $\alpha =0$.

\subsection{Stiff dynamics}
In the case of quintessential inflation after inflation the potential term quickly become 
subleading (i.e. $ \overline{a}^2 V \ll \varphi^{\,\prime\, 2}$) and, as already mentioned,
probably the simplest example along this direction is given by 
$V(\varphi) = \lambda (\varphi^4 + M^4)$ for $\varphi < 0$ and $V(\varphi) = \lambda M^8/(\varphi^4 + M^4)$ 
for $\varphi \geq 0$.  After inflation, for the quintessential potentials discussed here,  the effective potential 
$W$ coincides, in practice, with $V$; in other words we will have that $\overline{a}^2 W(\varphi) \simeq \overline{a}^2 V(\varphi)  \ll
 \overline{\varphi}^{\,\prime\, 2}$. If we consider the example given above the effective  potential becomes 
 \begin{equation}
 W(\varphi)  = \frac{\lambda M^4}{ [(\varphi/M)^4 + 1 + 8 \alpha \lambda (M/\overline{M}_{P})^2]} \simeq V(\varphi),
 \label{SS21}
 \end{equation}
 where the second equality follows since $\lambda \ll 1$ and assuming $M < \overline{M}_{P}$ and $\alpha \gg 1$. 
Values of $\alpha = {\mathcal O}(10^{2})$ 
seems to suffice to make the inflationary part of the potential compatible with current data \cite{anto1,enqv,anto2,tenka}. 
 
Thanks to Eq. (\ref{SS21}) during the stiff phase  $F\to 1$ and $\overline{F}_{1} 
\to 1$ so that the distinction between the Einstein and the Palatini frames disappears. 
This result can be easily discussed in the Palatini frame where it immediately follows from Eqs. (\ref{SS6}) 
and (\ref{SS7}). The solution of Eq. (\ref{SS6}) when the potential is negligible implies that 
$\varphi^{\prime} = \varphi_{1}^{\prime} (a_{1}/a)^2$. In the Palatini frame we also have that $F$ is 
given by Eq. (\ref{FF1b}) so that we obtain:
\begin{equation}
\overline{F} = 1 + 8 \, \ell_{P}^2 \, \overline{\alpha} \, V(\varphi) - 2\,  \overline{\alpha}\, \frac{\varphi^{\,\prime\, 2}}{a^2} \simeq 1 - 2 \alpha \,\biggl(\frac{H_{1}}{M}\biggr)^2 \, \biggl(\frac{a_{1}}{a} \biggr)^6.
\label{SS22}
\end{equation}
Equation (\ref{SS22}) implies that $F\to 1$ for $a> a_{1}$ so that, in this regime, 
the two frames coincide and the remaining equations together with Eq. (\ref{SS7}) can be solved:
\begin{equation}
a(\tau) \to \overline{a}(\tau) = \sqrt{\frac{\tau}{\tau_{1}}}, \qquad F \to 1.
\label{SS23}
\end{equation}
In the regime defined by Eqs. (\ref{SS22}) and (\ref{SS23}),  
$\overline{Q}$ will be suppressed. Indeed we will have that 
$\overline{Q} \, \simeq \alpha (H_{1}/M)^2 (\overline{a}_{1}/a)^6 \ll 1$ for $a > a_{1}$.
This observation is not only relevant for the evolution of the background but also for the evolution of the scalar and tensor inhomogeneities to be discussed hereunder. 

From the analysis of the inhomogeneities it will be established, among other things, that while the amplitude of the scalar power spectrum is not sensitive to the value of $\alpha$, the tensor-to-scalar ratio is suppressed in comparison with the case $\alpha =0$. In other words, anticipating the results of section (\ref{sec4}) (and in particular
of Eq. (\ref{QQ15})) we have that $r_{T}^{(\alpha)} = r_{T}^{(0)}/\sqrt{\overline{F}_{0}}$ while ${\mathcal A}_{{\mathcal R}}^{(0)}= {\mathcal A}_{{\mathcal R}}^{(\alpha)}= {\mathcal A}_{{\mathcal R}}$ where ${\mathcal A}_{{\mathcal R}}$ is simply the common value of the scalar power spectrum. Therefore, given a value of $r_{T}^{(\alpha)} \ll r_{T}^{(0)}$,  the expansion rate at the end of inflation is 
\begin{equation}
\frac{\overline{H}^{(\alpha)}}{M_{P}} = \frac{\sqrt{\pi\, {\mathcal A}_{{\mathcal R}} \, r^{(\alpha)}_{T}}}{4}, \qquad \frac{W}{M_{P}^4} = \frac{ 3\, r^{(\alpha)}_{T}\, {\mathcal A}^{(\alpha)}_{{\mathcal R}}}{128}.
\label{SS24}
\end{equation}
 By taking the fourth root of the second relation of Eq. (\ref{SS24}) and using the definition of the number of $e$-folds the following pair of relations can be obtained:
\begin{equation}
E^{(\alpha)} = \biggl(\frac{ 3\, r^{(\alpha)}_{T}\, {\mathcal A}_{{\mathcal R}}}{128}\biggr)^{1/4}, \qquad \biggl|\frac{\Delta \overline{\varphi}}{\Delta N} \biggr|= \overline{M}_{P} \sqrt{\frac{r^{(\alpha)}_{T}}{8}},
\label{SS25}
\end{equation}
where $E^{(\alpha)}$ is the typical energy scale of inflation and $|\Delta \overline{\varphi}/\Delta N|$ denotes the excursion of the inflaton field $\overline{\varphi}$ with the number of $e$-folds $N$. Moreover from  Eq. (\ref{SS24})  together with numerical values of $r^{(\alpha)}_{T}$ and ${\mathcal A}_{{\mathcal R}}$ the various scales 
can be written in more explicit terms: 
\begin{equation}
\frac{\overline{H}^{(\alpha)}}{M_{P}} = 2.17\times 10^{-6} \,\sqrt{\frac{r^{(\alpha)}_{T}}{0.01}}\,\, \sqrt{\frac{{\mathcal A}_{{\mathcal R}}}{2.4\times 10^{-9}}}.
\label{SS26}
\end{equation}
The presence of a post-inflationary stiff phase entails a modification of the maximal number of $e$-folds 
accessible to present observations. This effect can combine with the conservative reduction of the 
tensor-to-scalar ratio. Indeed the maximal number of inflationary $e$-folds 
accessible to large-scale CMB measurements can be derived, for the present purposes, 
by demanding that the inflationary event horizon redshifted at the present epoch coincides 
with the Hubble radius today:
\begin{equation}
e^{N^{(\alpha)}_{\mathrm{max}}} = \frac{[2 \pi\, \Omega_{R 0} \,{\mathcal A}_{{\mathcal R}}\,r^{(\alpha)}_{T}]^{1/4}}{4}\, \biggl(\frac{M_{P}}{H_{0}}\biggr)^{1/2}\, \biggl(\frac{\overline{H}^{(\alpha)}}{H_{r}} \biggr)^{1/2- \gamma},
\label{ST5}
\end{equation}
where $\Omega_{R 0}$ is the present energy density of radiation in critical units and $H_{0}^{-1}$ is the Hubble radius today. For the pivotal set of parameters of the concordance paradigm \cite{RT,RT2} Eq. (\ref{ST5}) becomes\footnote{In what follows $\Omega_{M0}$ and $\Omega_{R0}$ are the values of the critical fractions of matter and radiation in the concordance paradigm;  $h_{0}$ denotes the present value of the Hubble rate $H_{0}$ in units of $100 \,\mathrm{km}/(\mathrm{sec}\,\times\mathrm{Mpc})$.}:
 \begin{eqnarray}
 N^{(\alpha)}_{\mathrm{max}} &=& 60.74 + \frac{1}{4} \ln{\biggl(\frac{h_{0}^2 \Omega_{R 0}}{4.15 \times 10^{-5}} \biggr)} - \ln{\biggl(\frac{h_{0}}{0.7}\biggr)}
 \nonumber\\
 &+& \frac{1}{4} \ln{\biggl(\frac{{\mathcal A}_{{\mathcal R}}}{2.4 \times 10^{-9}}\biggr)} + \frac{1}{4} \ln{\biggl(\frac{r^{(\alpha)}_{T}}{0.01}\biggr)} + \biggl(\frac{1}{2} - \gamma\biggr) 
 \ln{\biggl(\overline{H}^{(\alpha)}/H_{r}\biggr)}.
 \label{ST6}
\end{eqnarray} 
In Eqs. (\ref{ST5}) and (\ref{ST6})  $\gamma$ (controlling the expansion in the stiff phase) accounts for the possibility of a delayed reheating terminating at a putative scale $H_{r}$ smaller (or even much smaller) than the Hubble rate during inflation.  

Since the reheating scale cannot be smaller than the one of nucleosynthesis, a lower bound 
on the possible extension of the stiff phase can be obtained by requiring 
 $H_{r}> 10^{-44} M_{\mathrm{P}}$ as it follows by demanding that the reheating occurs just prior to the formation of the light nuclei. When $\gamma > 1/2 $ (as it happens if $\gamma = 2/3$ when the post-inflationary background is dominated by dust), $N_{\mathrm{max}}$ diminishes in comparison with the sudden reheating (i.e. $\overline{H}^{(\alpha)}=H_{r}$) and $N_{\mathrm{max}}$ can become ${\mathcal O}(47)$.
Conversely if $\gamma <1/2 $ (as it happens in $\gamma = 1/3$ 
when the post-inflationary background is dominated by stiff sources ), $N^{(\alpha)}_{\mathrm{max}}$ increases.  
Finally, if $H_{r} = \overline{H}^{(\alpha)}$ (or, which is the same, if $\gamma=1/2$) there is a sudden transition 
between the inflationary regime and the post-inflationary epoch dominated by radiation. In spite of its dependence on ${\mathcal A}_{{\mathcal R}}$ and $r^{(\alpha)}_{T}$, the value of $N^{(\alpha)}_{\mathrm{max}}$ has then a theoretical error. Based on the previous considerations and on the maximal excursion of $\gamma$ we have that $N^{(\alpha)}_{\mathrm{max}} = 61.49 \pm  14.96$.

 If the total number of inflationary $e$-folds $N_{\mathrm{t}}$ 
exceeds $N^{(\alpha)}_{\mathrm{max}}$  the redshifted value of the inflationary event horizon is larger than the present value of the Hubble radius. If the inflationary piece of the total potential is quartic
the theoretical values of spectral index and  of the tensor-to-scalar ratio do not enter in the marginalized joint confidence contour in the plane $(n_{s},\, r_{T})$  at $2\sigma$  CL  \cite{RT,RT2} in the case $\alpha =0$. However, in the case 
of $\alpha = {\mathcal O}(10^{3})$ the quartic potential can be rescued \cite{anto1,enqv,anto2,tenka}.
Thanks to the larger number of $e$-folds typical of the stiff dynamics \cite{liddle,zao,mg3}
a complementary strategy could be to change the inflationary potential from quartic to quadratic as recently suggested \cite{haro}. This astuteness is however not necessary even if 
 the increase in the total number of $e$-folds and the suppression of the tensor-to-scalar ratio can still interfere constructively. 

In Eq. (\ref{ST6}) the scale $H_{r}$ has been used as a free parameter but its value can be fixed in a specific model. For instance, as orginally argued by Ford \cite{ford}, if ${\mathcal N}$ non-conformally coupled species are present during 
inflation their energy density might eventually dominate the stiff background. Since the energy density 
of ${\mathcal N}$ non-conformally coupled scalars is 
approximately of the order of $10^{-2} {\mathcal N} \overline{H}^{(\alpha)\,4}$, we could
estimate that, approximately, $H_{r} \simeq \overline{H}^{(\alpha)} (a_{1}/a_{r})^3 \sim 10^{-21} \, {\mathcal N}^{3/2} 
\, \overline{H}$ roughly corresponding to a temperature ${\mathcal O}(\mathrm{TeV})$. 
Further sources of radiation may be represented by the decay of massive particles \cite{haro}.  
The thermalization of the created quanta takes place quite 
rapidly also in the original case examined in Ref. \cite{ford}, and its specific occurrence is fixed by the moment 
at which the interaction rate becomes comparable with the Hubble expansion 
rate during the stiff phase \cite{ford,peevil,mg2}. While all these aspects 
are relevant, the explicit evolution of the inhomogeneities during and after 
inflation is even more crucial for the internal consistency of the whole discussion 
and this is why we cannot postpone further this analysis which will be 
developed in the following section.

\renewcommand{\theequation}{4.\arabic{equation}}
\setcounter{equation}{0}
\section{Scalar and tensor power spectra}
\label{sec4}
\subsection{Frame-invariance and gauge-invariance}
In  the Einstein frame the scalar and tensor modes of the geometry are:
\begin{eqnarray}
&& \delta_{s} \,\overline{g}_{00} = 2 \,\, \overline{a}^2\,  \overline{\phi},
\qquad \delta_{s} \,\overline{g}_{i j} 
= 2\,\, \overline{a}^2\, ( \overline{\psi}\,\, \delta_{ij} - \partial_{i}\partial_{j } \overline{E}),
\qquad \delta_{s}\, \overline{g}_{0i} = - \overline{a}^2 \,\, \partial_{i} \overline{B},
\label{TT1}\\
&& \delta_{t} \overline{g}_{i j} = - \overline{a}^2 \,\, \overline{h}_{ij},
\qquad  \partial_{i} \overline{h}^{ij} \, = \,\overline{h}_{i}^{i} =0,
\label{TT2}
\end{eqnarray}
where $\delta_{s}$ and $\delta_{t}$ denote, respectively, the scalar and tensor fluctuation of the corresponding quantity. 
The scalar and tensor modes of the geometry in the Palatini frame are given 
by the analog of Eqs. (\ref{TT1}) and (\ref{TT2}) with the difference that $\overline{a}$ is replaced 
by $a$ and the various fluctuations must be written without a bar; so for instance 
$\delta_{t} g_{ij} = - a^2 \, h_{ij}$,  $\delta_{s} \,g_{00} = 2 \, a^2 \, \phi$ and similarly for all the other 
perturbed entries of the metric.  

From the general expression of the conformal rescaling (i.e. $\overline{g}_{\mu\nu} = F g_{\mu\nu}$),
the expressions of the inhomogeneities in the two frames must be related as follows:
\begin{equation}
\delta_{s} \, \overline{g}_{\mu\nu} = (\delta_{s} F\,
g_{\mu\nu} + F \, \delta_{s} g_{\mu\nu}), \qquad \delta_{t} \overline{g}_{\mu\nu} = F \, 
\delta_{t} g_{\mu\nu}.
\label{TT2a}
\end{equation}
For the tensor modes the relation of Eq. (\ref{TT2a}) simplifies since 
$\delta_{t} F= 0$.  Thus the frame invariance of the tensor modes follows immediately from Eq. (\ref{TT2a}). Since $\delta_{t} \overline{g}_{\mu\nu} = F \,\delta_{t} g_{\mu\nu}$ we conclude that $\overline{a}^2(\tau) \, \overline{h}_{ij}(\vec{x},\tau) = 
F\, a^2(\tau)\, h_{ij}(\vec{x},\tau)$. But the background transforms as $\overline{a} = \sqrt{F} \, a$ and therefore we must conclude that $\overline{h}_{ij}(\vec{x},\tau) = h_{ij}(\vec{x},\tau)$. 
The tensor modes of the geometry are then separately invariant under infinitesimal diffeomorphisms in the two conformally related frames and they are therefore gauge-invariant and frame-invariant by construction.

The very same argument used in the case of the tensor modes  implies that only two metric fluctuations [out of the four appearing in Eq. (\ref{TT1})] are frame-invariant. More specifically, recalling that $\overline{a} = \sqrt{F} a$ we can write
\begin{equation}
\overline{\phi} = \phi + \delta q,\qquad \overline{\psi} = \psi - \delta q, \qquad \overline{E} = E, \qquad \overline{B} =  B,
\label{TT3}
\end{equation}
where $\delta q = \delta_{s} F/( 2 F)$, as it follows from the definition $q = \ln{\sqrt{F}}$. 
The gauge-invariant curvature fluctuation in the Einstein and Palatini frames are defined, respectively, as
\begin{equation}
\overline{{\mathcal R}} = - \overline{\psi} - \frac{\overline{{\mathcal H}}}{\overline{\varphi}^{\, \prime}}\chi ,\qquad 
{\mathcal R} = - \psi - \frac{{\mathcal H}}{\overline{\varphi}^{\,\prime}}\chi,
\label{TT4}
\end{equation}
where $\chi$ is the fluctuation of the scalar field while ${\mathcal H}$ and $\overline{\mathcal H}$ 
are related as in Eq. (\ref{SS5}):
\begin{equation}
\chi = \delta_{s}\overline{\varphi}  \equiv \frac{\delta_{s} \varphi}{\sqrt{\overline{F}_{0}}}, \qquad \overline{{\mathcal H}} = {\mathcal H} + q^{\prime}.
\label{TT4a}
\end{equation}
Since by definition $ \overline{\varphi}^{\,\prime} = \varphi^{\prime}/\sqrt{F_{0}}$ we also have that, in the curvature inhomogeneities, the contribution of the rescaled (scalar) fluctuations  is given by $\delta_{s}\,\overline{\varphi} (\overline{{\mathcal H}}/\overline{\varphi}^{\, \prime})= \delta_{s}\,\varphi (\overline{{\mathcal H}}/\varphi^{\, \prime})$. Thanks to the frame transformation (\ref{TT3}) we therefore conclude that $\overline{{\mathcal R}} = {\mathcal R}$ as long as $\delta q = (q^{\prime}/\varphi^{\prime}) \delta_{s} \varphi$. The curvature perturbations on comoving orthogonal hypersurfaces are then gauge-invariant 
and frame-invariant. Not all gauge-invariant fluctuations are also frame-invariant; for instance 
 the Bardeen potentials \cite{bardeen} are gauge-invariant but not frame-invariant. This is why 
 the description of the inhomogeneities in the longitudinal coordinate system is often not optimal 
 when dealing with the fluctuations of conformally related frames; this general conclusion 
 also applies to the present case.

\subsection{Evolution and relative normalization of the tensor modes}

In the two frames the evolution of the tensor modes reads, respectively  
\begin{equation}
\overline{h}_{i}^{j\,\,\prime\prime} + 2 \,\,\overline{{\mathcal H}}\,\, \overline{h}_{i}^{j\,\,\prime} - \nabla^2 \, \overline{h}_{i}^{j} =0, \qquad h_{i}^{j\,\,\prime\prime} + 2 ({\mathcal H} + q^{\prime}) \,\,h_{i}^{j\,\,\prime} - \nabla^2 h_{i}^{j} =0.
\label{TT5}
\end{equation}
But since the conformal rescaling of the background implies that
 $\overline{{\mathcal H}} = {\mathcal H} + q^{\prime}$ (see Eq. (\ref{SS5})) and since 
 according to Eq. (\ref{TT2a}), the two amplitudes must coincide (i.e. $h_{i\,j} = \overline{h}_{i\,j}$)
 we have to conclude that also the two equations appearing in Eq. (\ref{TT5}) coincide: this is a manifestation of the gauge-invariance and of the frame-invariance of the tensor modes. Thus mode expansion for the field operator corresponding 
to $\overline{h}_{i\, j}$ is given by:
\begin{equation}
\hat{\overline{h}}_{i\,j}(\vec{x}, \eta) = \frac{\sqrt{2} \ell_{P}}{(2\pi)^{3/2} \overline{a}(\tau)}\sum_{\lambda} \int \, d^{3} k \,\,e^{(\lambda)}_{ij}(\vec{k})\, \biggl[ f_{k,\lambda}(\tau) \hat{a}_{\vec{k}\,\lambda } e^{- i \vec{k} \cdot \vec{x}} + f^{*}_{k,\lambda}(\tau) \hat{a}_{\vec{k}\,\lambda }^{\dagger} e^{ i \vec{k} \cdot \vec{x}}\biggr],
\label{TT6a}
\end{equation}
where $e^{(\lambda)}_{ij}(\vec{k})$ (with $\lambda= \oplus,\, \otimes$) are the two tensor 
polarizations and the evolution of the mode functions follows from Eq. (\ref{TT5}):
\begin{equation}
f^{\prime\prime}_{k} + \biggl[ k^2 - \frac{\overline{a}^{\,\,\prime\prime}}{\overline{a}}\biggr] f_{k} =0,
\label{TT6b}
\end{equation}
where the explicit expression of $\overline{a}^{\prime\prime}/\overline{a}$ can be written in terms 
of the slow-roll parameter $\epsilon^{(\alpha)}$ already introduced in Eq. (\ref{SS16}):
\begin{equation}
\frac{\overline{a}^{\,\,\prime\prime}}{\overline{a}} =  \overline{a}^2 \, \overline{H}^2 [2 - \epsilon^{(\alpha)}] 
\equiv  \frac{\nu^{2} -1/4}{\tau^2}, \qquad \nu = \frac{ 3 - \epsilon^{(\alpha)}}{2 [ 1 - \epsilon^{(\alpha)}]}.
\label{TT6c}
\end{equation}
The tensor power spectrum follows by computing the expectation value of $\hat{\overline{h}}_{i\, j}$: 
\begin{equation}
\langle 0| \hat{\overline{h}}_{i\,j}(\vec{x},\tau) \hat{\overline{h}}_{i\, j}(\vec{y},\tau) |0\rangle = 
\int \,\,d\ln{k} \,\,{\mathcal P}^{(\alpha)}_{\mathrm{T}}(k,\tau) \,\, \frac{\sin{kr}}{kr},\qquad r = |\vec{x}- \vec{y}|,
\label{TT6d}
\end{equation}
where ${\mathcal P}^{(\alpha)}_{T}(k)$ is, by definition, the tensor power spectrum whose 
explicit form is:
\begin{equation}
{\mathcal P}^{(\alpha)}_{\mathrm{T}}(k) = \frac{4 \ell_{\mathrm{P}}^2 }{\overline{a}^2 \pi^2} k^3 |f_{k}(\tau)|^2 \equiv {\mathcal A}_{T}^{(\alpha)} \biggl(\frac{k}{k_{p}}\biggr)^{n_{T}^{(\alpha)}}.
\label{TT6e}
\end{equation}
While he first expression of Eq. (\ref{TT6e}) is general, the second equality defines the standard parametrization where 
$k_{p}$ denotes a coventional pivot wavenumber which should however coincide 
with the one employed in the parametrization of the scalar power spectrum [see also Eq. (\ref{Q11f}) and discussion therein]. 
The tensor spectral index and the corresponding amplitude are then defined as:
\begin{equation}
n_{T}^{(\alpha)} = 3 - 2 \nu \equiv - \frac{2 \epsilon^{(\alpha)}}{1 - \epsilon^{(\alpha)}} \simeq - 2 \epsilon^{(\alpha)} + {\mathcal O}(\epsilon^{(\alpha)\, 2})
\, \qquad {\mathcal A}_{T}^{(\alpha)} =\frac{128 \, W}{3 \,M_{P}^4},
\label{TT6f}
\end{equation}
where the exact definition of the tensor spectral index (see Eq. (\ref{TT6c})) has been included 
together with its slow-roll limit. If  $\alpha \to 0$ the tensor spectral index and the corresponding
 amplitude are:
\begin{equation}
n_{T}^{(0)} = - 2 \epsilon^{(0)}, \qquad {\mathcal A}_{T}^{(0)} = \frac{128\, V}{3\, M_{P}^4}.
\label{TT7}
\end{equation}
The mutual relation of Eq. (\ref{TT6d}) and Eq. (\ref{TT7}), to lowest order in the slow-roll approximation, is simply given by
\begin{equation}
n_{T}^{(\alpha)} = \frac{n_{T}^{(0)}}{\overline{F}_{0}}, \qquad  {\mathcal A}_{T}^{(\alpha)}= \frac{ {\mathcal A}_{T}^{(0)}}{\overline{F}_{0}},
\label{TT8}
\end{equation}
which is consistent with the expressions derived in Eqs. (\ref{SS15}) and (\ref{SS18}) by only 
considering the background evolution. 

\subsection{Evolution and relative normalization of the scalar modes}
 While the final evolution equation of  $\overline{{\mathcal R}}$ is both gauge-invariant and frame-invariant, the actual steps of the derivation may be more or less cumbersome depending 
on the specific gauge. We then suggest the gauge given by:
\begin{equation}
\delta_{s} \overline{g}_{00} = 2 \, \overline{a}^2 \, \overline{\phi}, \qquad \delta_{s} \overline{g}_{i0} = - \overline{a}^2 \, \partial_{i} \overline{B}, \qquad \delta_{s} \overline{\varphi} = \chi,
\label{QQ1}
\end{equation}
where the curvature perturbations on comoving orthogonal hypersurfaces coincide, in 
practice, with $\overline{\phi}$ [see hereunder Eq. (\ref{QQ8}) and discussion therein]. The perturbed version of Eq. (\ref{FF14}) can be 
written as $\delta_{s} \overline{G}_{\mu}^{\,\,\,\nu} = \ell_{P}^2 \delta_{s} T_{\mu}^{\,\,\,\nu}$
where $T_{\mu}^{\,\,\,\nu}$ denotes the effective energy-momentum tensor:
\begin{equation}
T_{\mu}^{\,\,\,\nu} = ( 1 + 2 Q ) \partial_{\mu} \overline{\varphi}\, \partial^{\nu} \overline{\varphi}  
+ \delta_{\mu}^{\nu} \biggl[  \frac{( 1 + Q)}{2} \overline{g}^{\alpha\beta} \partial_{\alpha} \overline{\varphi} \partial_{\beta} \overline{\varphi} - W\biggr],
\label{QQ1a}
\end{equation}
whose perturbed components, in the gauge (\ref{QQ1}), are\footnote{According to Eq. (\ref{SS14}), $\overline{Q}$ it is the homogeneous contribution of Eq. (\ref{FF14}).}:
\begin{eqnarray}
\delta_{s} T_{0}^{i} &=& - \frac{(1 + 2 \overline{Q})}{\overline{a}^2} \biggl[ \overline{\varphi}^{\, \prime} \partial^{i} \chi + 
\overline{\varphi}^{\, \prime\, 2} \partial^{i} \overline{B} \biggr],
\label{QQ2}\\
\delta_{s} T_{0}^{0} &=& \frac{1}{\overline{a}^2} \biggl[ (1 + 3 \overline{Q}) \biggl(\chi^{\prime} \, \overline{\varphi}^{\, \prime} - \overline{\varphi}^{\, \prime\, 2} \overline{\phi} \biggr) + \frac{3 \overline{\varphi}^{\, \prime\, 2} }{2}
\delta_{s} Q +  \frac{\partial W}{\partial \overline{\varphi}}  \overline{a}^2 \chi \biggr],
\label{QQ3}\\
\delta_{s} T_{i}^{j} &=&  \frac{\delta_{i}^{j}}{\overline{a}^2} \biggl[ (1 + \overline{Q}) \biggl(  \overline{\varphi}^{\, \prime\, 2} \overline{\phi} - \chi^{\prime} \, \overline{\varphi}^{\, \prime} \biggr) - \frac{\overline{\varphi}^{\, \prime\, 2} }{2}
\delta_{s} Q + \frac{\partial W}{\partial \overline{\varphi}}  \overline{a}^2 \chi \biggr].
\label{QQ4}
\end{eqnarray}
The explicit expression of $\delta_{s} Q$ appearing in Eqs. (\ref{QQ3}) and (\ref{QQ4}) follows from the scalar 
fluctuation of Eq. (\ref{FF14}) in the gauge (\ref{QQ1}) and it is:
\begin{equation}
\delta_{s} Q = 2 \, \overline{Q}\, \biggl[ \frac{\chi^{\prime}}{\overline{\varphi}^{\, \prime}} - \overline{\phi}\biggr] + 
8 \, \overline{Q}^2\,\, \frac{\overline{a}^2}{\overline{\varphi}^{\, \prime \, 2}}\,\, \frac{\partial W}{\partial \overline{\varphi}} \chi.
\label{QQ5}
\end{equation}
The explicit form of Eq. (\ref{QQ5}) has been derived by observing
that $F_{0}$ and $\delta_{s} F_{0}$ can be directly expressed in terms of  $W$ and of its fluctuations:
\begin{equation}
F_{0} = \frac{1}{1 - 8\, \overline{\alpha} \,\ell_{P}^2 \,W}, \qquad \delta_{s} F_{0} = \frac{8 \overline{\alpha} \,\ell_{P}^2 }{[1 - 8 \overline{\alpha} \ell_{P}^2 W]^2} \, \frac{\partial W}{\partial \overline{\varphi}} \, \chi.
\label{QQ5a}
\end{equation}
The exact evolution equations of the background (i.e. Eqs. (\ref{SS9})--(\ref{SS11}) and (\ref{SS13})) 
must be used to simplify the perturbed equations; we shall not dwell further on this aspect which 
should be however borne in mind throughout the remaining part of this section.
From the $(0i)$ component of the perturbed scalar equation (i.e. $\delta_{s} \overline{G}_{0}^{\,\,\,i} = \ell_{P}^2 \delta_{s} T_{0}^{\,\,\,i}$) we obtain the momentum constraint: 
\begin{equation}
2 \overline{{\mathcal H}} \,\, \overline{\phi} = \ell_{P}^2 \, \overline{\varphi}^{\, \prime} \,\chi \, (1 + 2 \overline{Q}).
\label{QQ6}
\end{equation}
Similarly, from the $(00)$ component of the perturbed scalar equation (i.e. $\delta_{s} \overline{G}_{0}^{\,\,\,0} = \ell_{P}^2 \delta_{s} T_{0}^{\,\,\,0}$) we obtain the Hamiltonian constraint  in the gauge (\ref{QQ1}):
\begin{equation}
2 \overline{{\mathcal H}} \,\nabla^2 \overline{B} + 6 \overline{\mathcal H}^2\, \overline{\phi} = 
- \ell_{P}^2\biggl\{ (1 + 3 \overline{Q}) \biggl[  \overline{\varphi}^{\,\prime} \chi^{\prime}- \varphi^{\, \prime\, 2} \overline{\phi} \biggr]+ \frac{\partial W}{\partial \overline{\varphi}}  \overline{a}^2 \chi + \frac{3}{2} \overline{\varphi}^{\, \prime 2} \delta_{s} Q \biggr\}.
\label{QQ7}
\end{equation}
In spite of the corrections (parametrized by $\overline{Q}$)  the relation of the curvature perturbations to the fluctuation of the scalar field is standard. Indeed, in the gauge 
(\ref{QQ1}) the expression of the curvature perturbations can be expressed as 
\begin{equation}
\overline{{\mathcal R}} = - \frac{\overline{\mathcal H}^2}{ \overline{\mathcal H}^2 - \overline{{\mathcal H}}^{\, \prime}} \, \overline{\phi} \equiv  - \frac{\overline{\mathcal H}}{\overline{\varphi}^{\, \prime}}\, \chi,
\label{QQ8}
\end{equation}
where the second equality follows exactly by eliminating $\overline{\phi}$ through Eq. (\ref{QQ6}) 
and by subsequently inserting Eq. (\ref{SS12}) into the obtained expression.  
The perturbed version of Eq. (\ref{FF15a})  can be finally rewritten, for the present purposes as: 
\begin{equation}
(1 + 2 Q) \overline{g}^{\alpha\beta} \overline{\nabla}_{\alpha} \overline{\nabla}_{\beta} \overline{\varphi} + 
2 \partial_{\alpha} Q \, \partial_{\beta} \overline{\varphi} \, \overline{g}^{\alpha\beta} + ( 1 - 4 Q^2) \frac{\partial W}{\partial \overline{\varphi}}=0.
\label{QQ1b}
\end{equation}
Since, according to Eq. (\ref{QQ8}), the relation between $\overline{{\mathcal R}}$ and $\chi$ 
is not affected by $\overline{Q}$ or by $\delta_{s} Q$, the evolution of $\overline{{\mathcal R}}$ follows by perturbing Eq. (\ref{QQ1b}); the result of this step is: 
\begin{eqnarray}
&& \chi^{\prime\prime} + 2 \,\biggl[ \overline{{\mathcal H}} + \frac{ \overline{Q}^{\, \prime}}{(2 \overline{Q} +1)}\biggr] \chi^{\prime} 
- \nabla^2 \chi + ( 1 - 2\, \overline{Q}) \frac{\partial^2 W}{\partial \overline{\varphi}^{\, 2}} \, \overline{a}^2 \, \chi \nonumber\\
&&+ 2 \, \overline{a}^2 \, \overline{\phi} \,( 1 - 2\, \overline{Q}) \frac{\partial W}{\partial \overline{\varphi}} 
- \overline{\varphi}^{\, \prime} ( \overline{\phi} + \nabla^2 \overline{B}) +
\frac{2\,\overline{\varphi}^{\prime}}{( 2\, \overline{Q} +1 )} \, \delta_{s} Q^{\prime} 
\nonumber\\
&& +  \frac{2 \delta_{s} Q}{( 2\, \overline{Q}+1)^2 } \biggl[ 2 \,\overline{Q}^{\, \prime} \, \overline{\varphi}^{\, \prime} + ( 2 \,\overline{Q}+ 1)^2 \frac{\partial W}{\partial \overline{\varphi}} \, \overline{a}^2 \biggr] =0.
\label{QQ9}
\end{eqnarray}
Inserting Eqs. (\ref{QQ6}) and (\ref{QQ7})  
into Eq. (\ref{QQ9}),  $\overline{\phi}$ and $\nabla^2 \overline{B}$ can be eliminated also from $\delta_{s} Q$ (see Eq. (\ref{QQ5})). Using finally Eq. (\ref{QQ8}) to express $\chi$ in terms of $\overline{{\mathcal R}}$ (i.e. $\chi = - \overline{\varphi}^{\, \prime} \overline{\mathcal R} /\overline{{\mathcal H}}$), Eq. (\ref{QQ9})
will give the explicit evolution of the gauge-invariant and frame-invariant curvature inhomogeneities:
 \begin{equation}
\biggl( \overline{\mathcal R}^{\,\prime} - \overline{\Delta}\biggr)^{\prime} 
+ \frac{[ \overline{z}^2 \, (1 + 2\, \overline{Q})]^{\,\prime}}{ \overline{z}^2 (1 + 2 \,\overline{Q})} \biggl( \overline{\mathcal R}^{\,\prime} - \overline{\Delta}\biggr) - \nabla^2 \overline{\mathcal R} =0, \qquad \overline{z} = \frac{\overline{a} \, \overline{\varphi}^{\, \prime}}{\overline{\mathcal H}},
\label{QQ10}
\end{equation}
where the expression of $\overline{\Delta}$ is:
\begin{eqnarray}
\overline{\Delta} &=& - \frac{4 \, \overline{Q}}{(1 + 2 \, \overline{Q})}  {\overline{\mathcal R}}^{\prime} + 
+ \frac{2 \, \overline{\mathcal R} }{(1 + 2\, \overline{Q})}
\biggl\{ \overline{Q}^{\, \prime} \biggl[ 1 + \frac{16\, \overline{Q}}{( 1- 4\, \overline{Q}^2)}\biggr] 
\nonumber\\
&+& 2 \, \overline{{\mathcal H}} \,\,\overline{Q} \biggl[ \biggl( \frac{\overline{\varphi}^{\prime}}{\overline{a}} \biggr)\biggl( \frac{\overline{a}}{\overline{\varphi}^{\prime}} \biggr)^{\prime} + \frac{4 \,\overline{Q}}{1 - 2 \,\overline{Q}} \frac{(\overline{a}^2 \overline{\varphi}^{\, \prime})^{\prime}}{(\overline{a}^2 \overline{\varphi}^{\, \prime})}\biggr]\biggr\}.
\label{QQ10a}
\end{eqnarray}
According to Eqs. (\ref{QQ10})--(\ref{QQ10a}), the same function controlling 
the evolution of the background also enters the evolution of curvature perturbations and this result 
guarantees the smallness of the corresponding corrections since, for opposite reasons, $\overline{Q} \ll 1$ both during the slow-roll evolution and during the stiff epoch.  In both regimes Eq. (\ref{QQ10}) can be expanded 
in powers of $\overline{Q}$ so that, to lowest order, the evolution of $\overline{{\mathcal R}}$ will then be given by:
\begin{equation}
\overline{{\mathcal R}}^{\,\prime\prime} + 2 \frac{\overline{z}^{\,\prime}}{z} \,\overline{{\mathcal R}}^{\,\prime} - \nabla^2 
\overline{{\mathcal R}} =0.
\label{QQ11}
\end{equation}
The mode expansion for the corresponding field operator $\hat{\overline{{\mathcal R}}}$ becomes:
\begin{equation}
\hat{\overline{{\mathcal R}}}(\vec{x}, \eta) = \frac{1}{\overline{z}(\tau) \, (2 \pi)^{3/2}} \int \, d^{3} k  \biggl[ g_{k}(\tau) \hat{b}_{\vec{k}} e^{- i \vec{k} \cdot \vec{x}} + g^{*}_{k}(\tau) \hat{b}_{\vec{k}}^{\dagger} e^{ i \vec{k} \cdot \vec{x}}\biggr], 
\label{Q11a}
\end{equation}
where the evolution of $g_{k}$, to lowest order in $\overline{Q}$, follows from Eq. (\ref{QQ11}) 
\begin{equation}
g^{\prime\prime}_{k} + \biggl[ k^2 - \frac{\overline{z}^{\,\,\prime\prime}}{\overline{z}}\biggr] g_{k} =0.
\label{Q11b}
\end{equation}
The explicit expression of $\overline{z}^{\prime\prime}/\overline{z}$ can be written in terms 
of $\epsilon^{(\alpha)}$ and $\eta^{(\alpha)}$ introduced in Eq. (\ref{SS16}) and the explicit result is:
\begin{eqnarray}
\frac{\overline{z}^{\,\,\prime\prime}}{\overline{z}} &=&  \overline{a}^2 \, \overline{H}^2 [ 2 + 5 \epsilon^{(\alpha)} - 9 \eta^{(\alpha)} 
+ (\eta^{(\alpha)} - \epsilon^{(\alpha)}) (\eta^{(\alpha)} - 2\epsilon^{(\alpha)}) ]
\nonumber\\
&\equiv&  \frac{\mu^{2} -1/4}{\tau^2}, \qquad \mu = \frac{ 3 + 3 \epsilon^{(\alpha)} - 2 \eta^{(\alpha)}}{2 [ 1 - \epsilon^{(\alpha)}]}.
\label{Q11d}
\end{eqnarray}
From the expectation value of $\hat{\overline{{\mathcal R}}}$ and with the same notations of Eq. (\ref{TT6d}): 
\begin{equation}
\langle 0| \hat{\overline{{\mathcal R}}}(\vec{x},\tau) \hat{\overline{{\mathcal R}}}(\vec{y},\tau) |0\rangle = 
\int \,\,d\ln{k} \,\,{\mathcal P}^{(\alpha)}_{{\mathcal R}}(k,\tau) \,\,\frac{\sin{kr}}{kr}, 
\label{Q11e}
\end{equation}
the scalar power spectrum ${\mathcal P}^{(\alpha)}_{{\mathcal R}}(k,\tau)$ reads, by definition, 
\begin{equation}
{\mathcal P}^{(\alpha)}_{{\mathcal R}}(k) = \frac{k^3}{2 \pi^2 \overline{z}^2}\,\, |g_{k}(\tau)|^2 \equiv {\mathcal A}_{{\mathcal R}}^{(\alpha)} \biggl(\frac{k}{k_{p}}\biggr)^{n_{s}^{(\alpha)}-1}, 
\label{Q11f}
\end{equation}
where $k_{p}$ denotes the same (but otherwise conventional) pivot scale used in Eq. (\ref{TT6e}). 
The scalar spectral index and the corresponding amplitude are:
\begin{eqnarray}
n_{s}^{(\alpha)} -1 = 3 - 2 \mu &\equiv& \frac{- 6 \epsilon^{(\alpha)} + 2 \eta^{(\alpha)}}{1 - \epsilon^{(\alpha)}} \simeq - 6\epsilon^{(\alpha)} + 2 \eta^{(\alpha)}+ {\mathcal O}(\epsilon^{(\alpha)\, 2}),
\nonumber\\
 {\mathcal A}_{{\mathcal R}}^{(\alpha)} &=&  \frac{8\, W}{3\, \epsilon^{(\alpha)} \, M_{P}^4}.
\label{Q11g}
\end{eqnarray}
As in Eq. (\ref{TT6f}) the exact definition of the tensor spectral index and its limit 
in the slow-roll approximation have been included.  When $\alpha =0$ the scalar spectral index and the corresponding 
spectral amplitude are given by:
\begin{equation}
n_{s}^{(0)} = 1 - 6 \epsilon^{(0)} + 2 \eta^{(0)}, \qquad {\mathcal A}_{{\mathcal R}}^{(0)} = \frac{8\, V}{3\, \epsilon^{(0)} \, M_{P}^4} .
\label{QQ12}
\end{equation}
Thanks to Eqs. (\ref{SS18}) and (\ref{SS19}) the scalar spectral indices of Eqs. (\ref{Q11g}) and (\ref{QQ12}) coincide,  
i.e. $n_{s}^{(0)} = n_{s}^{(\alpha)}$. The same conclusion holds 
for the corresponding spectral 
amplitudes: because of Eqs. (\ref{SS12}) and (\ref{SS18}) we have that $V/\epsilon^{(0)} = W/\epsilon^{(\alpha)}$ and therefore 
${\mathcal A}_{R}^{(0)}= {\mathcal A}_{R}^{(\alpha)}$. However, recalling Eqs. (\ref{TT7}) and (\ref{TT8}) the tensor to scalar 
ratio can be also computed and the result is 
\begin{equation}
r_{T}^{(0)} = \frac{ {\mathcal A}_{T}^{(0)}}{ {\mathcal A}_{{\mathcal R}}^{(0)}} = 16 \, \epsilon^{(0)} , \qquad 
r_{T}^{(\alpha)} = \frac{ {\mathcal A}_{T}^{(\alpha)}}{ {\mathcal A}_{{\mathcal R}}^{(\alpha)}} = 16 \, \epsilon^{(\alpha)}.
 \label{QQ14}
 \end{equation}
 Collecting together the results of Eqs. (\ref{SS12})--(\ref{SS18}) and (\ref{Q11g})--(\ref{QQ14})  we can therefore conclude that while the tensor-to-scalar ratio is suppressed the amplitude of the scalar power spectrum is not modified:
\begin{equation}
r_{T}^{(\alpha)} = \frac{r_{T}^{(0)}}{\sqrt{\overline{F}_{0}}}, \qquad {\mathcal A}_{{\mathcal R}}^{(0)}= {\mathcal A}_{{\mathcal R}}^{(\alpha)}= {\mathcal A}_{{\mathcal R}}
\label{QQ15}
\end{equation}
Equation (\ref{QQ15}) shows that 
the tensor-to-scalar ratio can be arbitrarily suppressed  by the presence of a term quadratic 
in the scalar curvature while the amplitude of the corresponding scalar power spectrum is left invariant and it is the same in both situations in spite of the value of $\alpha$.

\subsection{Spectrum of gravitational waves at high-frequencies}
When the observable modes are inside the Hubble radius (i.e. $k \tau \gg 1$), the spectral energy density (expressed in critical units) can be directly related to the tensor power spectrum 
(see, for instance, \cite{mg1,peevil,mg2,ford2}):
\begin{equation}
\Omega^{(\alpha)}_{gw}(k,\tau) = \frac{k^2}{12 a^2 H^2} {\mathcal P}^{(\alpha)}_{T}(k,\tau) \biggl[ 1 + {\mathcal O} \biggl(\frac{1}{k^2 \tau^2}\biggr) \biggr].
\label{GG1}
\end{equation}
The lowest frequency of the spectrum coincides, in practice, with the pivot scale appearing in Eqs. (\ref{TT6e}) and (\ref{Q11f}); choosing $k_{p} =0.002\,\mathrm{Mpc}^{-1}$ the corresponding pivot frequency is $\nu_{p} = k_{p}/(2 \pi) =  3.092\,\,\mathrm{aHz}$. In the conventional case the spectral energy density is quasi-flat for comoving frequencies $\nu$ ranging, approximately, between $100$ aHz and $100$ MHz. The transition across the epoch of matter-radiation equality leads to the low-frequency branch where 
$\Omega^{(\alpha)}_{gw}(\nu,\tau_{0}) \propto \nu^{-2}$ between the aHz and $100$ aHz, where 
$\tau_{0}$ denotes the present value of the conformal time coordinate.
In this regime the relevant transfer function can be expressed as:
\begin{equation}
T_{eq}(\nu,\nu_{eq}) = \sqrt{1 + c_{eq}\biggl(\frac{\nu_{\mathrm{eq}}}{\nu}\biggr) + b_{eq}\biggl(\frac{\nu_{\mathrm{eq}}}{\nu}\biggr)^2},
\label{RR5}
\end{equation}
where  $c_{eq}$ and $b_{eq}$ are two numerical constants of order $1$  and the typical 
frequency of the spectral transition is:
\begin{equation}
\nu_{eq}  =  1.362 \times 10^{-17} \biggl(\frac{h_{0}^2 \Omega_{\mathrm{M}0}}{0.1411}\biggr) \biggl(\frac{h_{0}^2 \Omega_{\mathrm{R}0}}{4.15 \times 10^{-5}}\biggr)^{-1/2}\,\, \mathrm{Hz}.
\label{RR6}
\end{equation}
As anticipated for $\nu \ll \nu_{eq}$ we have, approximately, 
$T^2_{eq}(\nu,\nu_{eq}) \simeq b_{eq} (\nu_{eq}/\nu)^2$. In the absence of a stiff post-inflationary phase, the suppression of $r_{T}^{(\alpha)}$ and of $\epsilon^{(\alpha)}$ 
in comparsion with $r^{(0)}_{T}$ and $\epsilon^{(0)}$ (see Eqs.  (\ref{SS15})--(\ref{SS18})  and (\ref{QQ14})) affects the Hubble rate at the end of inflation and therefore reduces the maximal frequency of the spectrum as: 
\begin{equation}
\nu^{(\alpha)}_{\mathrm{max}} = 1.95\times 10^{8} \biggl(\frac{\epsilon^{(\alpha)}}{0.001}\biggr)^{1/4} 
\biggl(\frac{{\mathcal A}_{{\mathcal R}}}{2.4\times 10^{-9}}\biggr)^{1/4} 
\biggl(\frac{h_{0}^2 \Omega_{R0}}{4.15 \times 10^{-5}}\biggr)^{1/4}  \,\, \mathrm{Hz},
\label{RR1}
\end{equation}
where, according to Eq. (\ref{QQ15}) ${\mathcal A}^{(\alpha)}_{{\mathcal R}}$ is the common
 amplitude of the scalar power spectrum which is unsuppressed even for $\alpha\neq 0$.
As $\epsilon^{(\alpha)}$ decreases the maximal frequency of the spectrum also decreases while the amplitude of the scalar fluctuations remains the same. However, since the tensor-to-scalar 
ratio is suppressed also the overall energy density of the gravitational waves 
will be suppressed in comparison with the case $\alpha =0$.  
The reduction implied by Eq. (\ref{RR1})  it is proportional to the quartic root 
of $\epsilon^{(\alpha)}$ and it is therefore not crucial. The normalizatization of the spectral energy density is instead 
reduced in a way proportional to $r_{T}^{(\alpha)}$ but this conclusion 
could be evaded, at least partially, in the case where $\alpha$ is scale 
dependent for instance because of an explicit coupling of the quadratic term 
to the inflaton.  This possibility will not be specifically discussed here. 

If the post-inflationary plasma is stiffer than radiation (and characterized by a generic barotropic index 
$w$ larger than $1/3$) the corresponding spectral energy density inherits a blue 
(or even violet) slope for typical frequencies larger than the mHz and smaller 
than about $100$ GHz (see, e.g. \cite{mg1,peevil,mg2,liddle,haro,mg5}). More precisely, in a model-independent 
approach, the frequency of the spike shall be expressed as 
\begin{equation}
 \nu^{(\alpha)}_{spike} = \nu^{(\alpha)}_{\mathrm{max}}/\sigma^{(\alpha)}, \qquad \sigma^{(\alpha)} =
\biggl(\frac{\overline{H}^{(\alpha)}}{H_{r}} \biggr)^{\frac{1-3w}{6 (w+1)}},
\label{RR1a}
\end{equation}
where $H_{r}$ denotes, as before, the Hubble rate at the epoch of radiation dominance.
According to Eq. (\ref{RR1a}) for a given $\overline{H}^{(\alpha)}$ the frequency of the spike is larger 
than $\nu^{(\alpha)}_{\mathrm{max}}$ provided $w > 1/3$. Moreover a decrease of   $\overline{H}^{(\alpha)}$ 
implies an increase of $\sigma^{(\alpha)}$ and a consequent decrease of the frequency of the spike. 
For $\alpha =0$ and $w=1$ the frequency of the spike typically reaches into the GHz band, given the values
of Eq. (\ref{RR1}). 

For the sake of simplicity we shall consider here the 
minimal situation where $\alpha$ is constant and scale-independent and, in this 
case, it can be shown that the energy density of the gravitational waves is given by
 \begin{equation}
h_{0}^2 \,\Omega^{(\alpha)}_{gw}(\nu,\tau_{0}) = {\mathcal N}_{\rho} \,\, r^{(\alpha)}_{T}( \nu_{p})\,\, {\mathcal T}^2(\nu, \nu_{eq},  \nu^{(\alpha)}_{s}) \, \,\biggl(\frac{\nu}{\nu_{\mathrm{p}}} \biggr)^{n^{(\alpha)}_{T}} \, e^{- 2 \,\beta\,\nu/\nu_{\mathrm{max}}}, 
\label{RR2}
\end{equation}
where $\beta={\mathcal O}(1)$ is a numerical parameter fixed that does not depend on 
$r_{T}^{(\alpha)}$; the overall normalization and the total transfer function can be expressed, respectively, as:
\begin{eqnarray}
{\mathcal N}_{\rho} &=& 4.165 \times 10^{-15}\, \biggl(\frac{h_{0}^2 \Omega_{\mathrm{R}0}}{4.15\times 10^{-5}}\biggr) \, \biggl(\frac{{\mathcal A}_{{\mathcal R}}}{2.41\times 10^{-9}}\biggr), 
\label{RR3}\\
{\mathcal T}(\nu, \nu_{eq},  \nu^{(\alpha)}_{s}) &=& T_{eq}(\nu,\nu_{eq}) \, T_{s}(\nu, \nu^{(\alpha)}_{s}).
\label{RR4}
\end{eqnarray}
where $T_{eq}(\nu,\nu_{eq})$ has been given in Eqs. (\ref{RR5}) and (\ref{RR6}). 
The parametrization of Eq. (\ref{RR4}) follows from the results of Ref \cite{mg5} and 
for $\nu \gg \nu_{eq}$ the second transfer 
function of Eq. (\ref{RR4}) (across $\nu^{(\alpha)}_{s}$) determines the high-frequency 
branch of the spectrum 
\begin{eqnarray}
T_{s}(\nu,\nu_{s}) &=& \sqrt{ 1 + c_{s}  \biggl(\frac{\nu}{\nu^{(\alpha)}_{s}}\biggr)^{p(w)/2} + b_{s}  \biggl(\frac{\nu}{\nu^{(\alpha)}_{s}}\biggr)^{p(w)}}, \qquad 
p(w) = 2 - \frac{4}{3w +1},
\label{RR7}\\
\nu^{(\alpha)}_{s} &=&  [\sigma^{(\alpha)}]^{3(w+1)/(3w -1)} \nu^{(\alpha)}_{\mathrm{max}}.
\label{RR8}
\end{eqnarray}
The values of $c_{s}$ and $b_{s}$ change depending on the values of $w$ (for $w\to 1$ there are even logarithmic corrections which have been specifically scrutinized in the past  but which are not essential here).  Since $\sigma^{(\alpha)}$ depends on $\overline{H}^{(\alpha)}$, a decrease in $r_{T}^{(\alpha)}$ amplifies  $\sigma^{(\alpha)}$. In this situation Eq. (\ref{RR8}) shows that a reduction of the frequency of the spike
is compensated by an increase of $\nu^{(\alpha)}_{s}$. The overall normalization  diminishes 
at small frequencies (i.e. in the aHz region) and the high-frequency branch of the spectral energy density 
gets squeezed implying a decrease of the signal  in the audio band (i.e. between few Hz and $10$ kHz). 
As previously suggested this conclusion may change if the inflaton and the higher-order gravitational corrections to 
the Einstein-Hilbert term are directly coupled.
\renewcommand{\theequation}{5.\arabic{equation}}
\setcounter{equation}{0}
\section{Concluding remarks}
\label{sec5}
Some of the actual realizations of the quintessential inflationary dynamics lead to a tensor-to-scalar ratio that is too large
and practically excluded by current data. Here we examined the possibility of complementing the gravitational action
 with contributions that are of higher-order in the Ricci scalar. Instead of considering a specific class of potentials, the impact of these contributions has been scrutinized  in the framework of  the Palatini formulation by positing that inflation already occurs in the conventional situation, i.e. when the gravitational action only contains the standard Einstein-Hilbert term. The analysis of the dynamics of the background and of the corresponding inhomogeneities shows that the addition of a quadratic term does not alter the scalar fluctuations but it just suppresses the tensor power spectrum and the tensor-to-scalar ratio. The initial value problems for the tensor and for the scalar modes of the geometry are well defined since the corresponding equations do not contain more than two derivatives with respect to the conformal time coordinate. However, unlike the evolution of the tensor modes, the equation of the gauge-invariant and frame-invariant curvature inhomogeneities on comoving orthogonal hypersurfaces is explicitly modified by the presence of higher-order corrections depending on the inflaton and on its derivatives. These potentially dangerous terms are everywhere negligible both during and after inflation so that, within the Palatini formulation, the power-law potentials leading to a quintessential inflationary dynamics are again viable. For a given spectral index the tensor-to-scalar ratio can always be adequately suppressed by the quadratic completion of the gravitational action. Moreover, for a sufficiently long stiff epoch, the maximal number of $e$-folds presently accessible by large-scale observations may get larger. The spectral energy density of the gravitons is still increasing at high-frequencies but the corresponding amplitude gets reduced; similarly the range of the high-frequency tail and the position of the spike are also reduced at least when the quadratic term is explicitly decoupled from the inflaton.
\section*{Acknowledgements}
It is a pleasure to thank T. Basaglia, A. Gentil-Beccot and S. Rohr of the CERN Scientific Information Service for their kind assistance.

\end{document}